# HOW FAR DOES A RANDOM FOREST GENERALIZE FROM A 54-RUN LAMMPS+SPICA BENCHMARK?

*Até onde uma Random Forest generaliza a partir de um benchmark LAMMPS+SPICA de 54 execuções?*

PEDERSEN, Dennis Alves[1]; RAMALHO, Paulo Henrique Leme[2]; ANDRIJAUSKAS, Fábio[3]
[1]Undergraduate Student in Software Engineering Bachelor Program at University of São Francisco, Itatiba, São Paulo, Brazil; [2]Master's Student in the Data Science in Health Program at University of São Francisco, Bragança Paulista, São Paulo, Brazil; [3]PhD Professor in Computer Engineering at University of São Francisco, Bragança Paulista, São Paulo, Brazil
dennis.pedersen@mail.usf.edu.br

**ABSTRACT.** Selecting near-optimal hybrid MPI+OpenMP configurations for molecular dynamics workloads on modern HPC clusters has traditionally required exhaustive empirical benchmarking, consuming allocation budget proportional to the number of configurations evaluated. This work investigates whether a cold-start Random Forest surrogate, trained once on a small, structured benchmark dataset, can reliably predict execution performance and recommend high-performing configurations without further cluster runs. The training dataset comprises 54 LAMMPS+SPICA runs of the antimicrobial peptide Tritrpticin on a hydrated DOPC bilayer (4 354 coarse-grained beads), spanning 18 hybrid configurations on 1–8 AMD EPYC 7662 nodes of the Lovelace cluster at CENAPAD-SP, with three independent replications each. Nine topology and resource features feed five regressors that predict loop time and four internal LAMMPS timing fractions (Pair, Kspace, Comm, Modify). In-sample mean absolute error is 0.49 s on loop time (4.0 % relative). Feature importance localizes predictive signal in topology variables (OpenMP threads and MPI/OpenMP ratio dominate; raw node and core counts contribute under 3 %). Leave-one-dimension-out generalization reveals that accuracy is governed by hardware regime membership: within a common regime (single-node, multi-node, or shared threading tier) the surrogate ranks configurations correctly, and degrades when targets cross architectural boundaries. The result is an interpretable map of where the surrogate's recommendations can be trusted, useful for scoping further benchmark campaigns at a fraction of their nominal cost.

**RESUMO.** A seleção de configurações híbridas MPI+OpenMP próximas do ótimo para cargas de dinâmica molecular em clusters HPC modernos tradicionalmente exige benchmarking empírico exaustivo, consumindo orçamento de alocação proporcional ao número de configurações avaliadas. Este trabalho investiga se um surrogate Random Forest treinado uma única vez ("cold-start") em um benchmark pequeno e estruturado pode prever com confiabilidade o desempenho de execução e recomendar configurações de alto desempenho sem novas execuções no cluster. O conjunto de treinamento reúne 54 execuções LAMMPS+SPICA do peptídeo antimicrobiano Tritrpticina em bicamada hidratada de DOPC (4 354 contas grosso-grão), em 18 configurações híbridas em 1–8 nós AMD EPYC 7662 do cluster Lovelace no CENAPAD-SP, com três replicações independentes cada. Nove atributos de topologia e recursos alimentam cinco regressores que predizem o loop time e quatro frações internas de tempo do LAMMPS (Pair, Kspace, Comm, Modify). O erro absoluto médio em amostra é 0,49 s no loop time (4,0 % relativo). A importância dos atributos localiza

o sinal preditivo em variáveis de topologia (threads OpenMP e a razão MPI/OpenMP dominam; contagens brutas de nós e núcleos contribuem com menos de 3 %). A generalização leave-one-dimension-out mostra que a acurácia é governada pela pertinência ao regime de hardware: dentro de um mesmo regime o surrogate classifica corretamente as configurações, e se degrada quando os alvos cruzam fronteiras arquiteturais. O resultado é um mapa interpretável de onde as recomendações do surrogate podem ser confiadas, útil para dimensionar campanhas de benchmark a uma fração do seu custo.

## INTRODUCTION

Selecting a near-optimal hybrid MPI+OpenMP partition for a molecular dynamics workload on a modern HPC cluster has remained, in practice, an exhaustive-search problem: each candidate configuration charges its full core-hour cost to the user's allocation whether or not it turns out to be competitive. As clusters grow in node count and intra-node architectural complexity, the configuration space expands combinatorially, and the cost of treating each new workload as a fresh empirical sweep becomes the dominant operational expense of pre-production tuning. The current state of the art for the MD case is exemplified by MDBenchmark (Gecht et al., 2020), a toolkit that automates the running of the full sweep — submission, collection, and statistical analysis of every candidate — rather than predicting performance from a subset, and whose authors explicitly frame the work in terms of the monetary, energetic, and environmental costs of MD simulations; its validation campaign measured 23 simulation systems across many MPI/OpenMP combinations precisely because no cheaper predictor was available.

The autotuning community has responded to the cost problem with surrogate-model-driven search: rather than benchmark every candidate, fit a regressor to a sparse sample of measurements and let the surrogate rank the rest. The ytopt framework (Wu et al., 2024) has demonstrated this on large-scale pre-exascale production systems, combining Bayesian optimization with a Random Forest surrogate to navigate search spaces of up to six million MPI/OpenMP configurations on Theta and Summit; the Kokkos-portable LAMMPS port of Johansson et al. (Johansson et al., 2025) delivers strong-scaling MD on Frontier, Aurora, and El Capitan, establishing the upper end of the resource regime where surrogate autotuning has been applied. Their common assumption, however, is that the user can afford an iterative tuning loop on the target cluster — hundreds to thousands of measurement points, Bayesian re-ranking, repeated evaluation. The opposite regime — a cold-start surrogate trained once on a small, structured benchmark and never updated — has received less attention, yet it is the regime that small research groups and short allocations actually live in.

This paper sketches a baseline for that regime. The training data is the 54-run LAMMPS+SPICA scalability dataset produced by our parent benchmarking study (Ramalho; Pedersen; Andrijauskas, 2026): a single coarse-grained workload (the antimicrobial peptide Tritrpticin on a hydrated DOPC bilayer), measured across 18 hybrid partitions on the AMD EPYC 7662 partition of CENAPAD-SP's Lovelace cluster. The surrogate is intentionally the cheapest defensible choice — a Random Forest regressor (Breiman, 2001) with off-the-shelf hyperparameters and no model selection — so that the results isolate what a minimum-effort surrogate can extract from a minimum-cost benchmark campaign. The deliberate choices to fix the workload, fix the architecture, and fix the model are scaffolding, not scope: each is a variable that subsequent work will relax.

Concretely, the practitioner's question is whether such a minimum-effort surrogate can be trusted to recommend a hybrid configuration one factorial step beyond what was actually measured — the next node-count doubling, the next OpenMP thread tier, or the next MPI-rank tier — so that a future tuning campaign can stop paying for benchmark runs that the surrogate could have predicted. Three pieces of information are needed before that question can even be posed. We first establish whether the surrogate has any signal at all by measuring its in-sample reconstruction of LAMMPS' internal timing breakdown across the 18-configuration sweep. We next identify which configuration features carry that signal — distinguishing raw resource quantities (nodes, total_cores) from topology variables (mpi_omp_ratio, OpenMP thread count) — so that the surrogate's strengths and blind spots are attributable to specific predictors and to identifiable model behaviour. With those two pieces in hand the per-axis generalization rates become interpretable in causal terms: each step-out outcome can be attributed to a specific configuration variable, so the rates tell us both whether the surrogate succeeds and the mechanism by which it succeeds, which in turn indicates where the failure modes will sit when the dataset is later expanded to other workloads or other architectures.

*Related work*

The works cited above sit at two different points on the cost-versus-coverage spectrum, and the present baseline targets a third. MDBenchmark (Gecht et al., 2020) automates exhaustive measurement rather than prediction, so its cost scales linearly with the number of candidate configurations regardless of how much structure those configurations share; it is the right tool when the full sweep itself is the deliverable, but it does not reduce the number of cluster-hours a new workload requires. ytopt (Wu et al., 2024) and the Kokkos-portable LAMMPS port (Johansson et al., 2025) sit at the opposite end: they assume repeated, cheap access to the target system, so that a Bayesian or evolutionary search loop can amortize its own measurement cost over hundreds to thousands of trials. Both regimes presuppose a measurement budget that a single short allocation does not have. The hybrid-MPI/OpenMP performance literature more broadly — exemplified here by the sparse-matrix-vector benchmarking of Schubert et al. (Schubert et al., 2011) — establishes why topology variables such as thread count and core-to-socket placement dominate runtime variance on ccNUMA hardware, which is the mechanistic explanation this paper's feature-importance result (see Results) is consistent with, but that literature does not address whether a regressor trained on one configuration set can be trusted to extrapolate to another. The present baseline is positioned in the gap between these strands: it asks how much predictive value a single, non-iterative, off-the-shelf surrogate can recover from a benchmark campaign that was never designed with model training in mind.

Answering this leading question on a single small workload establishes the floor of what a minimum-effort cold-start surrogate can extract, and motivates the follow-up campaign: training on multiple workloads simultaneously, transferring across force-field regimes, and replacing tree ensembles with extrapolation-aware learners. Those extensions are out of scope by design — we report the baseline first, and use its limits to scope what comes next.

## METHODOLOGY

*Benchmark dataset*

The surrogate is trained on the exact 54-row LAMMPS+SPICA scalability dataset produced by the parent benchmarking study (Ramalho; Pedersen; Andrijauskas, 2026) — the same table of loop_time_s measurements and internal LAMMPS timing fractions reported there, with no additional cluster runs of our own, archived unchanged in this repository as data/lammps_scalability.csv. It comprises 54 production runs spanning 18 distinct hybrid parallelization configurations (3 independent random seeds per configuration) executed on the AMD EPYC 7662 partition of the Lovelace cluster at CENAPAD-SP. The simulated system is the antimicrobial peptide Tritrpticin (PDB ID 1D6X) embedded in a hydrated DOPC bilayer, modelled with the SPICA coarse-grained force field (Seo; Shinoda, 2018) in LAMMPS (Thompson et al., 2022). The system contains 4 354 coarse-grained beads and combines Lennard-Jones short-range interactions with PPPM long-range electrostatics, a profile representative of electrostatically-active coarse-grained workloads.

Each configuration is described by the triple (nodes, np, omp) where np denotes the total number of MPI ranks and omp the OpenMP threads per rank, constrained so that np × omp = 128 × nodes (one full Lovelace node is 128 cores). Node counts span {1, 2, 4, 8}; configurations include pure-MPI (omp = 1), pure-OpenMP (np = nodes, omp = 128), and balanced hybrid points in between. The three independent seeds per configuration are kept to bound run-to-run variability; the empirical coefficient of variation is small (median 1.5 %, maximum 29 %), so the seed-averaged loop time is a reliable regression target.

For every run the dataset records LAMMPS' headline loop_time_s, derived throughput metrics (ns_per_day, timesteps_per_s), and the percentage of loop time spent in the internal sections Pair, Bond, Kspace, Neigh, Comm, Modify, Output, and Other as reported by LAMMPS' timing summary.

Table 1 – Training dataset: 18 hybrid MPI+OpenMP configurations measured at three seeds each on Lovelace (CENAPAD-SP, AMD EPYC 7662).

| nodes | np | omp | total_cores | seeds | loop_time_mean_s | loop_time_std_s |
|---|---|---|---|---|---|---|
| 1 | 1 | 128 | 128 | 3 | 65.575 | 0.154 |
| 1 | 8 | 16 | 128 | 3 | 11.827 | 0.022 |
| 1 | 16 | 8 | 128 | 3 | 8.047 | 0.119 |
| 1 | 32 | 4 | 128 | 3 | 5.763 | 0.271 |
| 1 | 64 | 2 | 128 | 3 | 4.649 | 0.068 |
| 1 | 128 | 1 | 128 | 3 | 4.398 | 0.113 |
| 2 | 32 | 8 | 256 | 3 | 9.111 | 0.1 |
| 2 | 64 | 4 | 256 | 3 | 11.523 | 0.244 |
| 2 | 128 | 2 | 256 | 3 | 10.007 | 0.391 |
| 2 | 256 | 1 | 256 | 3 | 16.688 | 0.111 |
| 4 | 64 | 8 | 512 | 3 | 8.287 | 0.156 |
| 4 | 128 | 4 | 512 | 3 | 9.442 | 0.103 |
| 4 | 256 | 2 | 512 | 3 | 10.354 | 0.12 |
| 4 | 512 | 1 | 512 | 3 | 14.563 | 0.098 |
| 8 | 128 | 8 | 1024 | 3 | 4.512 | 0.074 |
| 8 | 256 | 4 | 1024 | 3 | 5.978 | 0.213 |
| 8 | 512 | 2 | 1024 | 3 | 7.645 | 0.121 |
| 8 | 1024 | 1 | 1024 | 3 | 14.234 | 4.196 |

*Feature engineering*

From each row's (nodes, np, omp) triple we derive nine features that expose the configuration topology to the regressor in both linear and logarithmic encodings. The four raw-count features nodes, np, omp, and total_cores = np × omp describe the resource quantity directly. Two topology ratios — mpi_omp_ratio = np / omp and atoms_per_rank = N_beads / np — capture how those cores are partitioned and how much of the simulated system each rank carries. The remaining three features are base-2 logarithmic encodings of node, rank, and thread counts (log_nodes, log_np, log_omp); they make the strong-scaling regime appear approximately linear in feature space, which suits the axis-aligned splits used by tree ensembles (Breiman, 2001).

*Surrogate model*

We use Random Forest regressors (Breiman, 2001) as implemented in scikit-learn (200 trees, random_state = 42, default depth and split parameters). Five independent regressors are trained, one per target: the loop wall time (loop_time_s, seconds) and the four largest internal LAMMPS timing fractions (pair_pct, kspace_pct, comm_pct, modify_pct). Random Forests are used here as a one-shot regressor, rather than wrapped in an iterative Bayesian-optimization loop as in ytopt (Wu et al., 2024), for three reasons: they require no kernel selection on a small dataset; their tree-level variance yields an interpretable feature-importance ranking that maps directly onto MPI/OpenMP design decisions; and their inference cost is negligible relative to a single LAMMPS run.

Table 2 – Random Forest surrogate hyperparameters, feature set, and training targets.

| setting | value |
|---|---|
| estimator | sklearn.ensemble.RandomForestRegressor |
| n_estimators | 200 |
| random_state | 42 |
| max_depth | None (unrestricted) |
| min_samples_split | 2 (default) |
| min_samples_leaf | 1 (default) |
| bootstrap | True (default) |
| n_features | 9 |
| features | nodes, np, omp, total_cores, mpi_omp_ratio, log_np, log_omp, atoms_per_rank, log_nodes |
| n_targets | 5 |
| targets | loop_time_s, pair_pct, kspace_pct, comm_pct, modify_pct |
| training_rows | 54 |
| distinct_ configurations | 18 |
| seeds_per_ configuration | 3 |

*Reproducibility*

The full pipeline — feature derivation, model training, and the three evaluations below — is implemented in scripts/generalization_analysis.py against pandas>=2.0 and scikit-learn>=1.4 (pinned in requirements.txt); no feature scaling or normalization is applied, since tree ensembles split on raw feature values and are invariant to monotonic rescaling. Every leave-one-dimension-out subset trains an independent RandomForestRegressor instance from the same fixed random_state = 42 and the full n_estimators = 200, rather than reusing or warm-starting a single fitted model, so that each reported MAE and rank-of-best reflects only the training rows available to that subset. Held-out predictions are aggregated by (np, omp) and averaged across seeds before ranking, matching the seed-averaging already applied to the dataset-summary statistics in Table 1. The script and its CSV/figure outputs are checked into this repository, so every number reported in the Results section can be regenerated from data/lammps_scalability.csv without re-running LAMMPS.

*Evaluation protocol*

Three complementary evaluations, all reproducible from scripts/generalization_analysis.py, characterize the surrogate from different angles. The first is the in-sample fit: a regressor trained on the full 54-row dataset and evaluated on the same rows, which establishes the irreducible model error available given the chosen features and hyperparameters and serves as the lower bound against which any generalization measurement is read. The second is feature importance, computed as the mean decrease in impurity averaged across the 200 trees of the loop-time regressor and reported as relative contributions normalized to unity; this exposes which configuration variables carry the predictive signal. The third evaluation, on which the practical question of this paper hinges, is leave-one-dimension-out generalization. For each of the three configuration dimensions — nodes, np, and omp — we train on every non-empty subset of smaller values along that dimension and predict on the held-out larger value. Performance is measured both by mean absolute error in seconds and by the rank-of-best metric, defined as the rank that the surrogate assigns to the configuration whose seed-averaged loop time was actually lowest among the held-out configurations. A rank of 1 means the surrogate would have selected the truly optimal configuration without ever running it. Trivial cases, in which only a single held-out configuration remains and the rank is therefore degenerate, are filtered out before computing top-k rates.

## RESULTS AND DISCUSSION

*Predictive accuracy*

Across the full 54-row dataset the loop-time regressor achieves an in-sample mean absolute error of 0.49 s against a dataset-mean loop time of 12.4 s — a 4.0 % relative error. Per-configuration relative error is tight: the median is 2.3 %, 78 % of predictions fall within 5 % of the seed-averaged truth, and 94 % fall within 10 %. The 90th-percentile error is 6.9 %; only one configuration exceeds 15 % relative error on all three of its seeds — the 8-node, 1024-MPI-rank stress case, which also has the largest run-to-run variance in the dataset.

The four section-fraction regressors track their own targets with comparable fidelity (Table 3): MAE of 0.25, 0.67, 0.82 and 0.38 percentage points for Pair, Kspace, Comm and Modify respectively — between 1.7 % and 4.6 % of each section's dataset mean.

Table 3 – In-sample mean absolute error of each Random Forest regressor against the dataset mean.

| target | mae | dataset_mean | rel_error_pct |
|---|---|---|---|
| loop_time_s | 0.493 | 12.367 | 4.0 |
| pair_pct | 0.250 | 14.068 | 1.8 |
| kspace_pct | 0.670 | 39.234 | 1.7 |
| comm_pct | 0.823 | 31.199 | 2.6 |

| target | mae | dataset_mean | rel_error_pct |
|---|---|---|---|
| modify_pct | 0.381 | 8.192 | 4.6 |

*Feature importance*

Figure 1 reports the mean-decrease-in-impurity importances for the loop-time regressor. The four top-ranked features describe parallelization topology, not total resources: raw omp (22.4 %), mpi_omp_ratio (18.3 %), log_omp (18.1 %), and np (15.1 %) together account for 73.9 % of the model's explained variance. By contrast, nodes, total_cores, and log_nodes — the three features that encode raw resource quantity — together contribute only 2.9 % (all individually below 1.1 %). The system-size feature atoms_per_rank and log_np provide the remaining 23 %.

The practical implication is that how the cores are partitioned between MPI ranks and OpenMP threads matters far more than how many cores are used, in the resource regime spanned by this dataset. This is consistent with the hybrid-MPI/OpenMP literature (Schubert et al., 2011) which attributes thread-count sensitivity to ccNUMA locality-domain placement and intra-node communication patterns.

Figure 1 – Random Forest feature importance for the loop-time regressor. Topology features (omp, mpi_omp_ratio, log_omp, np) dominate; raw-resource features (nodes, total_cores, log_nodes) sit at the bottom.

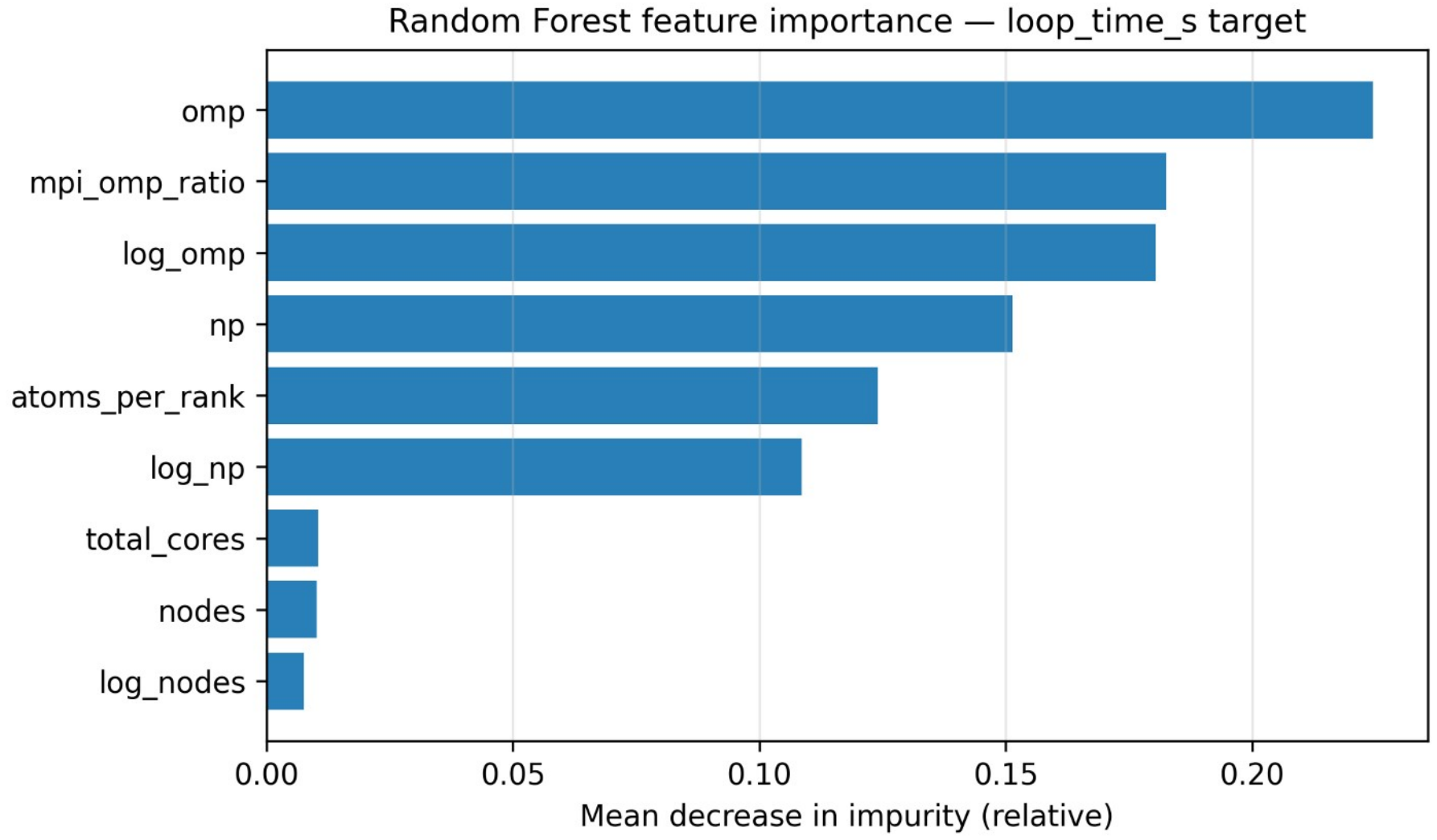

*Generalization across configuration dimensions*

Table 4 summarizes leave-one-dimension-out generalization on the loop-time regressor.

Along the OpenMP dimension the surrogate generalizes flawlessly to the moderately threaded regimes: when trained on configurations with omp ≤ 2 (and omp ≤ 4) it identifies the truly best configuration at the held-out omp = 2 and omp = 4 targets in 100 % of training subsets. At omp = 8 (one full CCX), generalization collapses to top-1 = 0 %, an extrapolation failure that we attribute below to the convex-hull limitation of tree ensembles.

Along the MPI-rank dimension the surrogate is reliable across most of the range: top-1 accuracy is 100 % at np = 32, 93 % at np = 64, and 96 % at np = 128. At np = 256 performance degrades sharply (top-1 = 22 %, top-2 = 41 %), partially recovering at np = 512 where top-1 = 59 % but top-2 = 100 % — the truly optimal configuration is always among the surrogate's top-two predictions when extrapolating to half-thousand-rank deployments.

Along the node dimension the surrogate correctly identifies the optimal configuration on 6 of 7 training subsets at the held-out nodes = 8 target (top-1 = 86 %) and on 2 of 3 subsets at nodes = 4 (top-1 = 67 %). The single nodes = 2 test case fails, but it is the most data-starved one in the study: only the 18 single-node rows are available for training, and none of the two-node (np, omp) configurations have an exact analogue at one node.

Table 4 – Leave-one-dimension-out generalization: top-1 and top-2 success rates for identifying the truly best held-out configuration.

| dimension | target | non_trivial_subsets | top1_pct | top2_pct |
|---|---|---|---|---|
| nodes | 2 | 1 | 0 | 0 |
| nodes | 4 | 3 | 67 | 67 |
| nodes | 8 | 7 | 86 | 86 |
| omp | 2 | 1 | 100 | 100 |
| omp | 4 | 3 | 100 | 100 |
| omp | 8 | 7 | 0 | 71 |
| np | 32 | 7 | 100 | 100 |
| np | 64 | 14 | 93 | 100 |
| np | 128 | 25 | 96 | 96 |
| np | 256 | 41 | 22 | 41 |
| np | 512 | 63 | 59 | 100 |

*Step-out probability along each configuration axis*

Aggregating the leave-one-dimension-out results as a function of the factorial step between the largest training value and the held-out target value isolates how the surrogate behaves at each doubling of the corresponding axis. Within the configurations sampled here, the top-1 success rate for predicting the optimal held-out configuration along the node axis is 67 % at the 4-node target and 86 % at the 8-node target, placing top-1 generalization to the next factorial step in node count at roughly four in five subsets. Along the OpenMP axis the surrogate stays fully reliable while the held-out thread count remains within a single CCX, identifying the optimum in 100 % of subsets at omp = 2 and again at omp = 4, while at omp = 8 it identifies the optimum in 0 % of subsets — a uniform failure at the first step that crosses the 8-thread CCX boundary. Along the MPI-rank axis the surrogate succeeds in 100 % of subsets at np = 32, 93 % at np = 64, and 96 % at np = 128, drops to 22 % at np = 256 (the first multi-node rank count in the dataset), and recovers to 59 % top-1 and 100 % top-2 at np = 512, placing the truly optimal configuration among the surrogate's two best guesses for every 512-rank subset.

The reading these rates support is conditional. Within the sampled hull the next factorial step in any single dimension is well predicted with probability at least 0.86 for nodes and at least 0.93 for moderate MPI rank counts. The surrogate's failure mode concentrates at the steps that cross an architectural boundary — one full CCX in OpenMP and the single-node to multi-node transition in MPI rank count — where successful extrapolation would require structural information about the hardware tier that the tree ensemble does not have access to through the engineered features.

*Limitations*

Three scope restrictions bound how far these results travel. First, the training data covers a single workload — Tritrpticin on a hydrated DOPC bilayer, 4 354 coarse-grained beads — and a single hardware generation (the AMD EPYC 7662 partition of Lovelace); nothing in the present dataset distinguishes a workload-specific effect from an architecture-specific one, since both are held fixed throughout. Second, the leave-one-dimension-out sweep draws on only 18 distinct configurations, so several held-out targets (notably nodes = 2 and omp = 8) are evaluated against a single non-trivial training subset; the reported top-1 rates at those targets are point estimates, not the law-of-large-numbers averages available at the better-sampled np targets. Third, the convex-hull failure mode identified here is diagnosed, not fixed: the Random Forest surrogate has no mechanism for flagging when a query configuration lies outside its training hull, so a deployed version of this baseline still requires the user to know in advance which architectural tier a candidate configuration belongs to.

Testing the first of these restrictions directly would mean benchmarking a much larger, intrinsically disordered membrane protein workload (on the order of 50× the bead count of the present system) on the same hardware, then applying this paper's Random Forest leave-one-out methodology to that larger workload. Whichever generalization geography emerges — the same convex-hull boundaries at the same architectural tiers, or a shifted geography driven by the larger system size — would indicate whether the failure modes catalogued here are a property of the surrogate or of this particular workload.

## CONCLUSIONS

The leading question admits a conditional yes. The in-sample fit is tight enough to confirm that the surrogate has signal to exploit; that signal lives almost entirely in topology features, with raw resource quantity contributing little. Read in light of those two pieces, the per-axis leave-one-dimension-out investigation provides a coherent geography of the surrogate's behaviour: a reliable interior in which the next factorial step remains inside an architectural tier and top-1 generalization holds at 86 % to 100 % of subsets, and attributable failures at the boundaries that cross those tiers — the one-full-CCX threshold for OpenMP threads and the single-node to multi-node transition for MPI rank counts. The proximate reason is the convex-hull limitation of tree ensembles, which restricts trustworthy prediction to the region of feature space the training subsets actually span. The practical value of the leave-one-out result is that the baseline returns a map of where its predictions can and cannot be trusted, which is more informative for scoping further measurement than a single aggregate accuracy figure on its own would be.

The baseline framing remains scaffolding for what comes next. Knowing the exact axes and step-out points at which the present surrogate breaks lets the natural follow-up campaign — multi-workload training to broaden the feature manifold, transfer across force-field regimes, and the replacement of tree ensembles with extrapolation-aware learners — be planned against concrete failure modes already documented in this paper. The most direct next step is to repeat the leave-one-dimension-out protocol on a substantially larger workload on the same hardware, which would indicate whether the convex-hull boundaries identified here are workload-specific or architecture-specific.

## ACKNOWLEDGMENTS

This work was carried out under the research project High-performance, high-throughput computing and machine learning techniques to create a framework for biological data processing, visualization, and sharing (CENAPAD-SP project proj1091), coordinated by Prof. Fábio Andrijauskas at Universidade São Francisco (USF). The benchmark dataset that this surrogate is trained on was produced under a CENAPAD-SP allocation on the Lovelace cluster (UNICAMP); the authors thank the CENAPAD-SP staff for that support and the LAMMPS and SPICA developer communities for the software stack.

During the preparation of this manuscript, the authors used large language model tools — Claude (Anthropic), the GPT series (OpenAI), and Gemini (Google) — for language editing, text drafting, literature search, and the development of analysis and automation scripts. The assistance was integrated into the workflow through the manuscript repository: a repository-resident agent ran sweeps of checks and tests against the data, and the continuous integration pipeline ran its own independent sweep on each pushed revision. All AI-assisted content was reviewed, critically evaluated, and validated by the authors, who take full responsibility for the integrity, accuracy, and originality of the work.

## DATA AND SOFTWARE AVAILABILITY

Training data: data/lammps_scalability.csv (this repository). Source code: this repository, commit-pinned in the final manuscript. Parent benchmarking dataset: (Ramalho; Pedersen; Andrijauskas, 2026).

## REFERENCES

BREIMAN, Leo. Random Forests. **Machine Learning**, v. 45, n. 1, p. 5–32, out. 2001.

GECHT, Michael *et al.* MDBenchmark: A toolkit to optimize the performance of molecular dynamics simulations. **The Journal of Chemical Physics**, v. 153, n. 14, out. 2020.

JOHANSSON, Anders *et al.* LAMMPS-KOKKOS: Performance Portable Molecular Dynamics Across Exascale Architectures. *In*: : SC Workshops '25.ACM, nov. 2025. Disponível em: <http://dx.doi.org/10.1145/3731599.3767498>

RAMALHO, Paulo Henrique Leme; PEDERSEN, Dennis Alves; ANDRIJAUSKAS, Fábio. **Strategies for Molecular Dynamics using Hybrid Systems: LAMMPS Use Case.**, jun. 2026. Disponível em: <https://arxiv.org/abs/2606.02319>

SCHUBERT, Gerald *et al.* Parallel Sparse Matrix-Vector Multiplication as a Test Case for Hybrid MPI+OpenMP Programming. *In*: IEEE, 2011. Disponível em: <http://dx.doi.org/10.1109/IPDPS.2011.332>

SEO, Sangjae; SHINODA, Wataru. SPICA Force Field for Lipid Membranes: Domain Formation Induced by Cholesterol. **Journal of Chemical Theory and Computation**, v. 15, n. 1, p. 762–774, dez. 2018.

THOMPSON, Aidan P. *et al.* LAMMPS - a flexible simulation tool for particle-based materials modeling at the atomic, meso, and continuum scales. **Computer Physics Communications**, v. 271, p. 108171, fev. 2022.

WU, Xingfu *et al.* ytopt: Autotuning Scientific Applications for Energy Efficiency at Large Scales. **Concurrency and Computation: Practice and Experience**, v. 37, n. 1, out. 2024.